\documentclass[11pt]{article}

\newcommand{\ba}{\begin{eqnarray}}
\newcommand{\ea}{\end{eqnarray}}

\newcommand{\SpSp}[2]{ \mbox{$\vec{\sigma }_{#1}.\vec{\sigma }_{#2}$}}
\newcommand{\lala}[2]{ \mbox{${\vec{\lambda}_{#1}\cdot
        \vec{\lambda}_{#2}}$}}

\begin{document}

\begin{center}
{\Large \textbf{Coloured s-wave quark clusters\\ with flavour
symmetry breaking}}
\vspace{25mm}

{\large H. H\o gaasen}\\
{\em Department of Physics
University of Oslo\\
Box 1048  NO-0316 Oslo Norway\\
hallstein.hogasen@fys.uio.no}\\

\vspace{10mm}

and\\

\vspace{10mm}

{\large P. Sorba} \\
{\em Laboratoire d'Annecy-le-Vieux de Physique Th\'{e}orique (LAPTH) \\
9 chemin de Bellevue, B.P. 110, F-74941 Annecy-le-Vieux Cedex,
FRANCE\\
sorba@lapp.in2p3.fr}

%\date{May 28., 2004}
%\maketitle

\end{center}

\vspace{20mm}

\begin{abstract}
We study the properties of coloured three particle s-wave  quark
clusters when flavour symmetry is broken. The relevance of such
clusters for  models of pentaquarks is shortly mentioned.

\end{abstract}

\vfill \vfill

\rightline{LAPTH-1074/04}
\rightline{hep-ph/0410224}
\rightline{sept. 2004}

\clearpage
\pagestyle{plain}
\baselineskip=18pt

\section{Introduction}
For the last thirty years colour magnetism has been a major tool
in understanding hadron spectroscopy.
It is still believed by many that the exchange of coloured gluons
provide the easiest way of explaining the mass differences between
hadronic states made up of the same (valence)
quarks \cite{dgg}.
When all spatial degrees of freedom are integrated out
we have an interaction Hamiltonian over colour spin space which is
the usual

\begin{equation}\label{1}
       H_{\mathrm{CM}} = - \sum_{i,j} C_{ij} \lala{i}{j} \SpSp{i}{j}
\end{equation}

\noindent Here the coefficients $C_{ij}$ are, among other things,
dependent
on the quark masses and properties of the spatial wave functions
of the quarks and the antiquark in the system. The solution of the
eigenvalue problem of the Hamiltonian above is therefore of
interest, not only in spectroscopy, but in all reactions where an
antiquark or a quark interact with a system of other quarks.\\
For this reason we think it is appropriate to present a full analysis
of
the colourmagnetic interaction with complete flavour symmetry breaking for
the $(qqq)$ and $(qq\overline{q})$ systems. It is a well defined problem of
undeniable interest, and we shall see that flavour symmetry breaking gives
effects
that are far more subtle than for the (trivial)
diquark $(q\overline{q})$ system.\\

\noindent The diagonalizing of $H_{\mathrm{CM}}$ in the cases where
baryons are made of three valence quarks and mesons of $(q\overline{q})$
    pairs, all coupled to a colour singlet, has been standard textbook
    stuff for a long time.
For hadrons with more complicated structure, where one
adds  extra $(q\overline{q})$ pair(s) to the simplest structures,
    new states ("multiquark states")are created, some of which
    carrying quantum numbers
that cannot be explained with
baryons made of three valence quarks and mesons made
of $(q\overline{q})$ pairs.
These are exotic states.

\noindent  After the recent reports of possible observations of
exotic baryons \cite{thetaexp}, the interest in
multiquark states has increased very much [3-13]. An extensive
list of references can be found in\cite{Zhu:2004xa}.\\
The search
for such states
had been strongly encouraged by D.Diakonov, using  predictions from the
chiral soliton model \cite{dpp}.
\noindent
  Studies of baryons with more than three quarks go back more than a quarter
of a century. At that time one made models where "coloured ions"
were bound together by colour-electric flux tubes \cite{Chan:1978nk} .
The mass defects
due to colour-magnetism were mostly made in the flavour symmetric
limit. A group theoretical mass formula was applied, the mass
defect being then  expressed in terms of the quadratic Casimir operators
for the $SU(2)$-spin, the $SU(3)$-colour and the $SU(6)$-colourspin group
\cite{jaffescalar}, \cite{Hogaasen:1978jw}. In
the cases where colour-spin, colour and spin for quarks and
     antiquarks can be
simultaneously quantized together with the same operators
     for the whole system,the results are quite easily generalized
to flavour symmetry breaking. In other cases not.

\noindent As can be immediately noticed by elementary
group theory computations, some multiquark states are
{\em mixtures} of states of colour
$SU(3)_c$, spin $SU(2)_s$ and colour spin $SU(6)_{cs}$
representations, and that implies some care in their treatment.

\noindent Such a situation appears as soon as a set of quarks and
antiquarks is considered, and these are among the cases of s-wave
clusters
that we address in this paper.

\noindent From the decomposition of $SU(3)$ representations:

\begin{equation}\label{2}
       3 \times 3\times 3 = 1 + 8 + 8 + 10  \end{equation} and
       \begin{equation}
       3  \times 3 \times \overline{3} = 3 + \overline{6} + 3 + 15
\end{equation}

\noindent one deduces that the possible s-wave triplets of quarks belong
to
one of the following colour representations $c = 1, 8, 10$, while
clusters made of $2q$ and $1\overline{q}$ stand in $c = 3, 6$ or $15$, the
total spin being for each cluster $s= 1/2$ or $s= 3/2$.

\noindent Let us add that quark triplets in colour octet $(qqq)^8_{s}$ as
well as $(qq \overline{q})^{\overline{6}}_{s}$ configurations have a long
   time ago been considered in
some detail, mostly in the flavour symmetry limit, as part
of exotic baryons. Such "pentaquarks", earlier called
"meso-baryonium" \cite{Hogaasen:1978jw}, were constituted by two
clusters of the type:
$(qqq)^8 - (q\overline{q})^8$  or $(qq\overline{q})^{\overline{6}}
- (qq)^6$, each cluster standing in an s-wave, and separated from
the other by an orbital angular momentum barrier. The
configuration $(qq\overline{q})^3_{1/2} -
(qq)^{\overline{3}}_{0}$, again with a relative L=1, has also been
recently used \cite{kl1} to make pentaquark states of spin $1/2^{+}$: a
detailed analysis of this approach, based from a study of
$H_{\mathrm{CM}}$ with flavour symmetry breaking, is performed in
\cite{hogso}.

\noindent We present our results over a set of base states.
   We first couple two
quarks $q_1$ and $q_2$ into a definite colour and spin system, then
add a third quark/antiquark and consider the resulting clusters with
definite colour and spin. Note that different orderings of quarks in
   a cluster with specified colour $c$ and spin $s$ generally lead
to different Hamiltonian matrices. For example, choosing $u$ and $d$
    as quarks $q_{1}$ and $q_{2}$ in the $(udc)$ system will provide an
    Hamiltonian matrix different from the one obtained by choosing the
    quarks $d$ and $c$ for the same positions 1 and 2. The eigenvalues
    will of course be  the same, as well as the physical content
    of the eigenvectors.\\

\noindent How one numbers the quarks is not quite
    without importance however:
the reduction of the Hamiltonian obtained by imposing some
(approximate) flavour
symmetry, such as isospin symmetry, is explicit if u and d are chosen as quarks
1 and 2.
In the isospin symmetry case  we have noted that our calculations are
in agreement, as
could be expected,  with earlier results in the three flavour sector
\cite{Chan:1980cj}.
For quarks with identical flavour, the effect of the Pauli exclusion
principle is most easily incorporated by considering them as $q_{1}$
and $q_{2}$.\\
When we sometimes refer to SU(3) flavour representations it evidently does not
mean that we consider u,d and s quarks only,the number 3 comes just
   because we study three particle systems, so that the (maximal) number
   of different flavours in a cluster is 3.\\
   The eigenvectors of  $H_{\mathrm{CM}}$ are usually not falling into one
   specific representation of $SU(3)_{f}$, they are naturally
   (also called magically) mixed combinations of different  $SU(3)_{f}$
   representations.

\section{ The $(qqq)^8_{s}$ triquark}

This type of triquark had shown up in one of the two favorite
configurations - the second one implying the $(qq\overline{q})^6_{s}$
cluster - proposed in  \cite{Hogaasen:1978jw} for narrow exotic
baryons made of two
clusters protected one form the other by a relative angular momentum
barrier.

\noindent Let us denote $q_{1}$,$q_{2}$ and $q_{3}$ the three quarks under
study, all in a relative s-wave.

\vspace{0.5cm}

\noindent i) We consider first the case of total spin $s = 1/2$, that is $
|q_1 q_2 q_{3}\rangle^{8}_{1/2}$.

\noindent Reminding the following
decompositions of $SU(3)$ representations:

$$3 \times 3 = \overline{3} + 6 \ \ \  ;\ \ \   \overline{3}\times 3 = 1 +
8 \ \ \ ; \ \ \ 6 \times 3 = 8 + 10, $$ it follows that the space
on which the Hamiltonian $H_{\mathrm{CM}}$ acts is four
dimensional and a natural basis is provided with the four states:
\begin{eqnarray}\label{4}
       \chi_1 &=& |(q_1 q_2)^6_1 \rangle \otimes |
       (q_3)^{3}_{1/2} \rangle \nonumber \\
       \chi_2 &=& |(q_1 q_2)^{\overline{3}}_1 \rangle \otimes |
       (q_3)^{3}_{1/2} \rangle \nonumber \\
           \chi_3 &=& |(q_1 q_2)^6_0 \rangle \otimes |
       (q_3)^{3}_{1/2} \rangle \nonumber \\
           \chi_4 &=& |(q_1 q_2)^{\overline{3}}_0 \rangle \otimes |
       (q_3)^{3}_{1/2} \rangle
\end{eqnarray}
the notation here being:
\begin{eqnarray}\label{4}
       \chi_1 &=& |(q_1 q_2)^c_s \rangle \otimes |
       (q_3)^{3}_{1/2} \rangle
       \end{eqnarray}

\noindent here $c$ is the colour and $s$ the spin of the doublet of the
two
quarks $q_{1}$ and $q_{2}$.

\noindent The states  $\chi_1$  and $\chi_4$  have the pair of two quarks
$(q_{1}q_{2})$which are
coupled symmetrically in colour-spin and are therefore belonging
to the $(6 \times 6)_S = 21$ dimensional representation of
$SU(6)_{cs}$, the states $\chi_2$  and $\chi_3$ are antisymmetric
in colour-spin of the two quarks $(q_{1}q_{2})$ and fall in the $(6 \times
6)_A =
15$ dimensional representation.

\noindent Note that if the two quarks $q_{1}$ and $q_{2}$ are identical in
flavour, the states
$\chi_1$ and $\chi_4$ vanish due to the Pauli principle.

\noindent A way to explicitly compute the $4 \times 4$ matrix representing
$H_{CM}$ relative to the $(qqq)^8_{1/2}$ triplet is to
study separately the colour part and the spin part namely:
\begin{equation}\label{5}
       H_C = - \sum_{i,j} C_{ij} \overrightarrow{\lambda}_i
       \cdot \overrightarrow{\lambda}_j \ \ \ \ \ H_S = -  \sum_{i,j} C_{ij}
\overrightarrow{\sigma}_i
       \cdot \overrightarrow{\sigma}_j
\end{equation}
and then to perform a kind of "tensor product" of the two
so-obtained $2 \times 2$ matrices.

\noindent Let us consider the colour-action part. Then, when acting by
$H_C$
on the $\chi_i$'s, it will be convenient to express $| (q_1 q_2)^c
(q_{3})^{3} \rangle^8$ where $c =6$ or
$\overline{3}$ in terms of  $| (q_1q_3)^c (q_2)^{3}
\rangle^8$ and $| (q_2q_3)^c (q_1)^{3} \rangle^8$ (we omit the lower
spin index in this computation). By direct calculation, one obtains
the colour crossing:

\begin{eqnarray}\label{6}
       V_c \equiv
       \left(
\begin{array}{c}
| (q_1 q_2)^6 (q_3)^3 \rangle^8 \\
| (q_1 q_2)^{\overline{3}}(q_{3})^{3}
\rangle^8
\end{array}
        \right) & = &
        \left[
        \begin{array}{cc}
        -\frac{1}{2} & \sqrt{\frac{3}{2}} \\
         - \sqrt{\frac{3}{2}} &  -\frac{1}{2}
        \end{array}
        \right] \ \
        \left(
\begin{array}{c}
| (q_2 q_3)^6 (q_1)^{3} \rangle^8 \\
| (q_2 q_3)^{\overline{3}} (q_1)^{3} \rangle^8
\end{array}
        \right) \nonumber \\
         & = &
         \left[
        \begin{array}{cc}
        -\frac{1}{2} & \sqrt{\frac{3}{2}} \\
        \ \   \sqrt{\frac{3}{2}}  & \ \  \frac{1}{2}
        \end{array}
        \right] \ \
        \left(
\begin{array}{c}
| (q_1 q_3)^6 (q_2)^{3} \rangle^8 \\
| (q_1 q_3)^{\overline{3}} (q_2)^{3} \rangle^8
\end{array}
        \right)
\end{eqnarray}
from where we can also derive the (inverse) expressions of $| (q_1
q_3)^c (q_2)^{3} \rangle^8$ and $| (q_2
q_3)^c (q_1)^{3} \rangle^8$ in terms of $| (q_1 q_2)^c
(q_3)^3 \rangle^8$. It is then
straightforward to derive the $H_C$ matrix.

\noindent A similar technics will allow to construct the $2 \times 2 \
H_S$
matrix, and we finally give the complete expression for the
colour magnetic Hamiltonian $H_{CM}$ acting on the 4-dim vector
$\overrightarrow{\chi} = (\chi_1, \chi_2, \chi_3, \chi_4)$:\\
$H_{CM}$ =
\begin{equation}\label{7}
    - \left[
\begin {array}{cccc}
\frac{4}{3}\,{\it C_{12}}+\frac{10}{3} ({\it C_{13}} +{\it
C_{23}}) & 2 \sqrt {3} \left( {\it C_{13}}-{\it C_{23}} \right) &
\frac{5}{\sqrt {3}} \left( {\it C_{13}}-{\it C_{23}} \right)
&3
     \left( {\it C_{13}}+{\it C_{23}} \right) \\\noalign{\medskip}
     2 \sqrt {3} \left( {\it C_{13}}-{\it C_{23}} \right) & -\frac{8}{3}\,
     {\it C_{12}}-\frac{2}{3}\,({
\it C_{13}} + {\it C_{23}})& 3 \left( {\it C_{13}}+{\it
C_{23}} \right) & -\frac{1}{\sqrt{3}} \left( {\it C_{13}}-{\it
C_{23}} \right) \\\noalign{\medskip} \frac{5}{\sqrt {3}} \left(
{\it C_{13}}-{\it C_{23}}  \right) &3 \left( {\it
C_{13}}+{\it C_{23}} \right) &-4\,{\it
C_{12}}&0\\\noalign{\medskip} 3 \left( {\it
C_{13}}+{\it C_{23}} \right) & -\frac{1}{\sqrt {3}} \left( {\it
C_{13}}-{\it C_{23}} \right) &0&8\,{ \it C_{12}}
\end {array}
\right]
\end{equation}
\vspace{4mm}

\noindent In the flavour symmetry limit, then
$C_{12}$ = $C_{23}$ = $C_{31}$ = $C$
and $H_{CM}$ reduces to:

\begin{equation}\label{8}
H_{\mathrm{CM}} = -C \cdot \left[ \begin {array}{cccc} 8&0&0&6
\\\noalign{\medskip}0&-4 &6&0\\\noalign{\medskip}0&6
&-4&0\\\noalign{\medskip}6&0&0&8\end {array}
     \right]
\end{equation}

\vspace{4mm}

\noindent

\noindent
and the space state decomposes into two invariant subspaces. One
is spanned by $\chi_1$  and $\chi_4 $, its $SU3$ flavour content is a
singlet and an octet with eigenvalues -14C and -2C
    for $H_{\mathrm{CM}}$. The other, spanned by
$ \chi_2 $ and $\chi_3 $, contains a flavour  octet and a decuplet
    where the eigenvalues of $ H_{\mathrm{CM}}$ are -2C and +10C
respectively.

\noindent Let us note that the determination of the associated $SU(3)$
flavour
    representations is naturally obtained from the corresponding
    colour-spin $SU(6)_{cs}$ ones. Indeed, in $SU(6)$:

    \begin{equation}\label{2}
       6 \times 6 = 21 + 15  \end{equation}
    with the $SU(3)_{c}\times SU(2)_{s}$ decompositions:

    $$6 = (3, 1/2) \ \ \   ;\ \ \    21 = (6, 1) + (\overline{3}, 0)\ \ \
;\ \ \    15 =
    (6, 0) + (\overline{3}, 3)$$
(in which the $SU(3)$ part is denoted by its dimension and the
    $SU(2)$ part by its $s$-label), and we remark that the couples
    ($q_{1}q_{2}$) in $\chi_1$  and $\chi_4 $ belong to the symmetric
    21 representation of $SU(6)_{cs}$, while in $\chi_2$  and $\chi_3 $
    they stand in the antisymmetric 15 one. Moreover:

     \begin{equation}\label{2}
      21 \times 6 = 56 + 70
    \ \ \   {\mbox{and}} \ \ \  15 \times 6 = 20 + 70  \end{equation}
    Then corresponding $SU(3)$ multiplets are selected by insuring the
    complete antisymmetry of  $SU(6)_{cs}\times SU(3)_{f}$.
    As noted before, if two quarks are identical in flavour it is convenient
to label them as
    particles 1 and 2 so that the states $\chi_1$  and $\chi_4 $ are
forbidden
    by the Pauli principle.\\

\noindent In this case, and in all other cases where we have a system of
only
quarks (or only antiquarks)  the eigenstates in the flavour symmetric
limit (only!) correspond
to sharp values of the
    total colour-spin. Examining the $SU(3)_{c}\times SU(2)_{s}$
    decompositions:
$$ 56 = (8, 1/2) + (10, 3/2)   $$ $$  70 = (8, 1/2 + 3/2) + (10 +1,
1/2)   $$ $$ 20= (8, 1/2) + (1, 3/2)$$

\noindent we note that the eigenvectors which are mixtures of our base
states
are identical to the 56 and 70
    representations as well as of the 70 and 20 representations of
    $SU(6)_{cs}$, any of these $SU(6)$ representations containing an
    $SU(3)$ colour octet and an $SU(2)$ spin doublet.\\
    When flavour symmetry is broken this is no longer the case, and
$H_{\mathrm{CM}}$ and the total
    colour-spin can no longer be simultaneously quantized.

    \noindent Finally, it is interesting to remark that , if the three
    particles are identical, then only the combination ( $\chi_2$ -
    $\chi_3$) is allowed by the Pauli principle, as can be seen hereafter.
    Indeed the flavour sector for the three quarks must then be the
    symmetric 10 dim. representation of $SU(3)_{f}$, implying
    corresponding states to belong to the completely antisymmetric 20
    dim. representation of  $SU(6)_{cs}$. As explicited just above, such
    a configuration involves for the doublet ($q_{1}q_{2}$) the 15 of
$SU(6)_{cs}$
    and the product $15\times 6$ ( see Eq(11) ). But a direct
    Clebsch-Gordan computation can show that the combination ( $\chi_2$ -
    $\chi_3$) belongs exactly to the 20 of $SU(6)_{cs}$ while ( $\chi_2$ +
    $\chi_3$) belongs to the 70 one.
    \noindent An indirect check of this result can be obtained from the
    $ H_{\mathrm{CM}}$ matrix, which immediately reduces to a $2\times2$
    matrix, $\chi_1$  and $\chi_4 $ being forbidden as already remarked:

    \begin{equation}\label{9}
H_{\mathrm{CM}}=-\left[ \begin {array}{cc} -4\,{\it C} &  6{\it C}
\\\noalign{\medskip}  6{\it C}&
{-4\it C}\end{array} \right].
\end{equation}

    \noindent Diagonalizing $ H_{\mathrm{CM}}$ is immediate and provides
    the two eigenvalues $10C$ and $-2C$ corresponding to the eigenvectors
( $\chi_2$ - $\chi_3$) and ( $\chi_2$ + $\chi_3$) respectively.
Now, referring to our old computation of \cite{Hogaasen:1978jw} where, in the
symmetry limit case, expectation value of $ H_{\mathrm{CM}}$ have
been computed for the three quark cluster $\Theta_{f}($c,s$)$, we
recognize the $H$ eigenvalue $10C$ for $\Theta_{10}(8,2)$ with
$\Theta_{10}$ transforming under the 20 of $SU(6)_{cs}$ and $-2C$
for $\Theta_{8}(8,2)$ with $\Theta_{8}$ transforming under the 70
of $SU(6)_{cs}$ (appearing twice).\\

\noindent ii) The case  $|q_1 q_2 q_{3}\rangle^{8}_{3/2}$ is
    simpler to study since there is no "spin mixing", and only the
    colour part  of $ H_{\mathrm{CM}}$ is not trivial,
    that is:

     \begin{equation}\label{9}
H_{\mathrm{CM}}=-\left[ \begin {array}{cc} \frac{4}{3}\,{\it
C_{12}}-\frac{5}{3} ({\it C_{13}} +{\it
C_{23}}) & - \sqrt {3} \left( {\it C_{13}}-{\it C_{23}} \right)
\\\noalign{\medskip}
    -\sqrt {3} \left( {\it C_{13}}-{\it C_{23}} \right) & -\frac{8}{3}\,
     {\it C_{12}}+\frac{1}{3}\,({
\it C_{13}} + {\it C_{23}}) \end {array} \right]
\end{equation}

\noindent acting on the two dimensional space spanned by $\chi_1$
and $\chi_2$.

\section{ The $(qqq)^c_{s}$ clusters with colour $c=10$ and
$c=1$.}

\noindent These are examples in which the calculation of the colour
magnetic Hamiltonian is very simple. For example, in order to form a
colour decuplet - resp. a colour singlet - of three quarks, any couple of
quarks must be in a sextet - resp. antitriplet - representation of
$SU(3)_{c}$. A direct computation gives for $\sum_{i,j} \lala{i}{j}$
the value:
\begin{equation}
       K_{6} = 4/3
\end{equation}
for the sextet and
\begin{equation}
       K_{\overline{3}} = - 8/3\end{equation} for the
(anti)triplet.

\noindent When the total spin of the cluster is $s = 1/2$, then over
the basis:
\begin{eqnarray}
\pi_1 &=& |(q_1 q_2)_1^{c^\prime} \rangle \otimes |
(q_3)^{3}_{1/2} \rangle^c \nonumber \\
\pi_2 &=& |(q_1 q_2)_0^{c^\prime} \rangle \otimes |
(q_3)^{3}_{1/2} \rangle^c \nonumber \\
\end{eqnarray}
\noindent with $c= 10$ and $c' = 6$ or $c = 1$ and $c' =
\overline{3}$, one finds , using the spin crossing matrix:

\begin{equation}\label{9}
H_{\mathrm{CM}}=- K_{c} \left[ \begin {array}{cc}{\it
C_{12}}- 2 ({\it C_{13}} +{\it
C_{23}}) & - \sqrt {3} \left( {\it C_{13}}-{\it C_{23}} \right)
\\\noalign{\medskip}
    -\sqrt {3} \left( {\it C_{13}}-{\it C_{23}} \right) & -3
     {\it C_{12}}
    \end {array}\right ]
\end{equation}

\noindent with $K_{6} = 4/3$ for the $(qqq)^{10}_{1/2}$ cluster and
$K_{\overline{3}} = -8/3$ for the usual baryons  $(qqq)^1_{1/2}$.\\
The Pauli principle forbids the state $\pi_2$ ($\pi_1$) if two quarks
    chosen as $q_1$ and $q_2$
have identical flavour and the cluster is a colour singlet (colour
decuplet).
If the flavor is identical for all three quarks, then both  $\pi_2$  and$\pi_1$
are forbidden.

\noindent
If the total spin is $s = 3/2$, then there is only one
state, i.e. $\pi^c_{1}$, and $H_{\mathrm{CM}}$ reduces to :

\begin{equation}\label{9}
H_{\mathrm{CM}}= - K_{c} ( C_{12} + C_{23} + C_{13})\end{equation}

\noindent

    \noindent Three quark clusters in colour decuplet belong to $SU(3)$
    flavour octets when $s=1/2$ and $SU(3)$ flavour singlets when
    $s=3/2$. Similarly, three quark clusters in colour singlet are
    naturally connected to flavour octets when $s=1/2$ and to flavour
    decuplets when $s=3/2$.

\noindent  In the flavour symmetry limit, one gets as
$H_{\mathrm{CM}}$ eigenvalues $+ 4C$ (resp.$- 4C$) for $(qqq)^8_{1/2}$
(resp.$(qqq)^8_{3/2}$) clusters), and $+ 8C$ (resp. $-8C$) for
$(qqq)^1_{1/2}$
(resp.$(qqq)^1_{3/2}$) clusters.

\section{ The $(qq\overline{q})^3_{s}$ triquark}

    This  type of
triquark state, in the case s = 1/2, has recently been used
\cite{kl1} , together
with a
spin zero diquark state carrying colour $\bar{3}$, to make
pentaquark states of spin ${1/2}^+$ when the triquark and diquark
are separated by a L=1 orbital angular momentum.
A detailed study of the flavour symmetry breaking has been already
performed in our paper \cite{hogso}. Therefore we will
rapidly provide
hereafter with the colour-magnetic Hamiltonian matrix, spending
however sometime discussing the limit cases.
We consider first the case of total spin s = 1/2. The two quarks
$q_1$  and $q_2$ can be coupled to colour  $\bar{\mathbf{3}}$ or
$\mathbf{6}$, to spin 0 or spin 1. Together with the antiquark
$\bar{q}_3$, spin and colour couplings are such that the cluster
carries total colour 3 and spin 1/2. It follows that  the space on which
the Hamiltonian (Eq.\ref{1})
acts over is four dimensional and a natural basis is provided with
the four states:
\begin{eqnarray}\label{2}
       \phi_1 &=& |(q_1 q_2)^6_1 \rangle \otimes |
       (\overline{q}_3)^{\overline{3}}_{1/2} \rangle \nonumber \\
       \phi_2 &=& |(q_1 q_2)^{\overline{3}}_1 \rangle \otimes |
       (\overline{q}_3)^{\overline{3}}_{1/2} \rangle \nonumber \\
           \phi_3 &=& |(q_1 q_2)^6_0 \rangle \otimes |
       (\overline{q}_3)^{\overline{3}}_{1/2} \rangle \nonumber \\
           \phi_4 &=& |(q_1 q_2)^{\overline{3}}_0 \rangle \otimes |
       (\overline{q}_3)^{\overline{3}}_{1/2} \rangle
\end{eqnarray}

\noindent
For completeness, let us remind
the following product decompositions of $SU(3)$ representations:
\begin{equation}\label{4}
       3 \times 3 = \overline{3} + 6 \ \ \ \ ; \ \ \ \ \overline{3}
       \times \overline{3} = 3 + \overline{6} \ \ \ \ ; \ \ \ \ 6
       \times \overline{3} = 3 + 15
\end{equation}
The states  $\phi_1$  and $\phi_4$  have  two quarks which are
coupled symmetrically in colour-spin and are therefore belonging
to the $(6 \times 6)_S = 21$ dimensional representation of
$SU(6)_{cs}$, the states $\phi_2$  and $\phi_3$ are antisymmetric
in colour-spin of the two quarks and fall in the $(6 \times 6)_A =
15$ dimensional representation.

\noindent Note that if the two quarks are identical in flavour, the states
$\phi_1$ and $\phi_4$ vanish due to the Pauli principle.

\noindent The complete expression for the
colour magnetic Hamiltonian $H_{CM}$ acting on the 4-dim vector
$\overrightarrow{\phi} = (\phi_1, \phi_2, \phi_3, \phi_4)$ reads :\\
$H_{CM}$ =
\begin{equation}\label{7}
    - \left[
\begin {array}{cccc}
\frac{4}{3}\,{\it C_{12}}+\frac{20}{3} ({\it C_{13}} +{\it
C_{23}}) & 4 \sqrt {2} \left( {\it C_{13}}-{\it C_{23}} \right) &
\frac{10}{\sqrt {3}} \left( {\it C_{13}}-{\it C_{23}} \right)
&2\sqrt {6}
     \left( {\it C_{13}}+{\it C_{23}} \right) \\\noalign{\medskip}
     4 \sqrt {2} \left( {\it C_{13}}-{\it C_{23}} \right) &
-\frac{8}{3}\,{\it C_{12}}+\frac{8}{3}\,({
\it C_{13}} + {\it C_{23}})& 2 \sqrt{6} \left( {\it C_{13}}+{\it
C_{23}} \right) & \frac{4}{\sqrt{3}} \left( {\it C_{13}}-{\it
C_{23}} \right) \\\noalign{\medskip} \frac{10}{\sqrt {3}} \left(
{\it C_{13}}-{\it C_{23}}  \right) &2\sqrt {6} \left( {\it
C_{13}}+{\it C_{23}} \right) &-4\,{\it
C_{12}}&0\\\noalign{\medskip} 2\, \sqrt{6} \left( {\it
C_{13}}+{\it C_{23}} \right) & \frac{4}{\sqrt {3}} \left( {\it
C_{13}}-{\it C_{23}} \right) &0&8\,{ \it C_{12}}
\end {array}
\right]
\end{equation}
\vspace{4mm}

\noindent It is easily seen from this matrix that, if we impose flavour
symmetry for the two quarks ($C_{12}$ = $C_{23}$), we get a matrix
operating over two invariant subspaces $\{ \phi_1 , \phi_4 \}$ and $\{
\phi_2 , \phi_3 \}$ respectively.\\
If, in addition we impose full flavour symmetry for the
     interaction and assume that the $qq$ and $q\bar{q}$ interactions are the
same
     (so that $C_{ij}=C$), then we have the matrix:\\
\begin{equation}\label{8}
H_{\mathrm{CM}} = -C \cdot  \left[ \begin {array}{cccc} {\frac
{44}{3}}&0&0&4\,\sqrt {6}
\\\noalign{\medskip}0& \frac{8}{3} &4\,\sqrt
{6}&0\\\noalign{\medskip}0&4\,
\sqrt {6}&-4&0\\\noalign{\medskip}4\,\sqrt {6}&0&0&8\end {array}
     \right]
\end{equation}

\vspace{4mm}

\noindent and we fall back on the old results
\cite{Chan:1978nk}, \cite{Mulders:1978cp}  where the
eigenvalues of the colourmagnetic interaction are - 21.88C and
- 0.98C for the case when the two quarks are coupled symmetrically
     in colour-spin. For antisymmetric colour spin the eigenvalues are
     - 9.68C and + 11.02C.

\noindent In no case are the eigenvectors corresponding to sharp
     values of the total colour-spin. They are mixtures of the 6
     and 120 dimensional representations as well as of the 6 and
$\overline{84}$
     representations of the colour spin $SU(6)_{cs}$ algebra when
     considering the $(qq\overline{q})^3_{1/2}$ system. Indeed,
     performing the product of $SU(6)$ representations:
     \begin{equation}\label{9}
       21 \times \overline{6} = 6 + 120 \ \ \ \ \mbox{and} \ \ \ \ 15
       \times \overline{6} = 6 + \overline{84}
\end{equation}
and examining the corresponding $SU(3) \times SU(2)$
decompositions
\begin{equation}\label{10}
       6 = (3 \ , \frac{1}{2}) \ \ \ \ \ \ \ \  120= (3+15 \ ,  \frac{1}{2} +
       \frac{3}{2}) + (\overline{6} \ , \frac{1}{2}) \ \ \ \ \ \ \ \
\overline{84}
       = (15 \ , \frac{1}{2}) + (3+\overline{6} \ , \frac{1}{2} +
       \frac{3}{2})
\end{equation}
one easily remarks that both the 6 and 120 $SU(6)$ representations
contain a triplet of colour and doublet of spin, and that is also
the case for the couple of representations 6 and $\overline{84}$.

\noindent Moreover, if we decouple the antiquark (going to the heavy
quark limit
or
     considering
     relative spatial wave functions that have no $s$-wave overlap)
     putting $C_{13}$ = $C_{23}=0$, the effective Hamiltonian $H_{CM}$ is
     diagonal, with elements
     which are the well known colour magnetic energies for colour  sextet and
triplet diquarks.

\noindent As has been remarked before, if the two quarks are identical in
     flavour, the matrix is $2 \times 2$ and the states $\phi_1$ and
     $\phi_4$ disappear.

   \noindent  In the flavour symmetry limit, the states $\phi_1$ and
     $\phi_4$ which have the two quarks  in the symmetric colour spin
     representation 21 are associated with the flavour $SU(3)$
     representation $f = \overline{3}$, while the states $\phi_2$ and
     $\phi_3$ stand in the $f = 6$ representation as the two quarks are
     in the antisymmetric representation of colour spin.

\noindent Note that the flavour content $(qq\overline{q})$ is
     $\overline{3}\times \overline{3} = 3 + \overline{6}$ for $\phi_1$ and
     $\phi_4$ and $6 \times \overline{3} = 3 + 15$ for $\phi_2$ and
     $\phi_3$.

\noindent When the triquark $(qq\overline{q})$ is combined with the
     (most strongly bound) diquark $(qq)$ which has
     $c=\overline{3}, s=0$ and flavour $f = \overline{3}$, the total
     $(qqqq\overline{q})$ states containing $\phi_1$ and
     $\phi_4$ will be in the flavour representation $(3+\overline{6}
     ) \times \overline{3} = 1+8+8+\overline{10}$, while the states
     containing $\phi_3$ and $\phi_4$ will be in the  $(3+15) \times
     \overline{3}= 1+8+8+10+27$ flavour representations.

   \noindent The representations $\overline{10}$ in the first group, and 27 in
     the second group, manifestly contain exotics.

   \noindent As we have seen, $\phi_1$ and
     $\phi_4$ will mix as well as $\phi_2$ and
     $\phi_3$ if there is colour magnetic interaction
     $(C_{q\overline{q}} \neq 0)$ between the antiquark and the
     quarks. When flavour symmetry is broken, all states will in
     general mix: this corresponds to mixing of states in different
     flavour representations.

   \noindent If we use isospin symmetric $u$ and
     $d$ quarks, then $C_{13} = C_{23}$ and some states with different
     flavour symmetry will not mix. This is the case for all models
     of the exotic $\Theta^+$ which is assumed to be $(ud \ ud\overline{s})$,
     and it will only belong to the $ f =\overline{10}$
     representation.\\

\noindent We conclude by the total spin 3/2 case. Then the matrix
representation of
$H_{cm}$
      is acting over the space:

\begin{eqnarray}\label{8}
       \pi_1 &=& |(q_1 q_2)^{6}_1 \rangle \otimes |
       (\overline{q}_3)^{\overline{3}}_{1/2} \rangle \nonumber \\
       \pi_2 &=& |(q_1 q_2)^{\overline{3}}_1 \rangle \otimes |
       (\overline{q}_3)^{\overline{3}}_{1/2} \rangle \nonumber \\
\end{eqnarray}

and reads:
\vspace{5mm}
     \begin{equation}\label{9}
H_{\mathrm{CM}}=-\left[ \begin {array}{cc} \frac{4}{3}\,{\it
C_{12}}-\frac{10}{3} ({\it C_{13}} +{\it
C_{23}}) & - 2 \sqrt {3} \left( {\it C_{13}}-{\it C_{23}} \right)
\\\noalign{\medskip}
    - 2 \sqrt {3} \left( {\it C_{13}}-{\it C_{23}} \right) & -\frac{8}{3}\,
     {\it C_{12}}-\frac{4}{3}\,({
\it C_{13}} + {\it C_{23}}) \end {array} \right]
\end{equation}

\vspace{14mm}
\section{ The $(qq\overline{q})^{\overline{6}}_{s}$ triquark}

This is the second type of triquark (the first one being the colour
octet one) to which we devoted a special attention
\cite{Hogaasen:1978jw} for
constructing possibly narrow multiquark baryons made of two clusters.
    In contrast with the previous case where the cluster carries
    colour $ \overline{3}$, we have here a system where there is
    $\underline{no}$ combination
     of $(q\overline{q})$
    that are invariant under $SU(3)_c$ transformations. Evidently
    all $(q\overline{q})$
    must be in colour octets in order to couple with a colour triplet and
    provide a colour  $ \overline{6}$.
    This triquark (and the ones in the following sections) is therefore
     protected from dissociation into a quark and a colour singlet meson.\\
\noindent The space over which $H_{\mathrm{CM}}$ acts is two dimensional, and
    as basic states we choose:
\begin{eqnarray}\label{8}
       \psi_1 &=& |(q_1 q_2)^{\overline{3}}_1 \rangle \otimes |
       (\overline{q}_3)^{\overline{3}}_{1/2} \rangle \nonumber \\
       \psi_2 &=& |(q_1 q_2)^{\overline{3}}_0 \rangle \otimes |
       (\overline{q}_3)^{\overline{3}}_{1/2} \rangle \nonumber \\
\end{eqnarray}
\noindent It is understood here that here colour = $\overline{6}$  and spin is
$1/2$ for the $(qq\overline{q})$ system.\\
    It should be noted that we have again chosen a basis where states
    are not eigenstates for total colour-spin and that the state $\psi_2$ is
    not present if the two quarks have identical flavour.
    For the state $\psi_1 $, colour-spin is a mixture of 6 and $\overline{84} $
    (with flavour 3 and 15 respectively), while for the state $\psi_2 $,
    the colour-spin is a mixture of 6 and 120 ( with flavour 3 and
$\overline{6} $).\\
\noindent Over this basis the Hamiltonian reads:

    \begin{equation}\label{9}
H_{\mathrm{CM}}=-\left[ \begin {array}{cc} -\frac{8}{3}\,{\it
C_{12}}-\frac{4}{3} ({\it C_{13}} +{\it
C_{23}}) & - 2 \sqrt {3} \left( {\it C_{13}}-{\it C_{23}} \right)
\\\noalign{\medskip}
    - 2 \sqrt {3} \left( {\it C_{13}}-{\it C_{23}} \right) & 8\,
     {\it C_{12}} \end {array} \right]
\end{equation}\\

\noindent In the flavour symmetric case the Hamiltonian is diagonal.

\noindent When this triquark $(qq\overline{q})$ is combined with the
     (most strongly bound) diquark $(qq)$ which has
     $c=\overline{6}, s=1$ and flavour $f =\overline{3}$, the total
     $(qqqq\overline{q})$ states containing $\psi_1$  stand in the
     flavour representation $(3+15
     ) \times \overline{3} = 1+8+10+27$, while the states
     containing $\psi_2$ stand in the  $(3+ \overline{6}) \times
     \overline{3}= 1+8+8+ \overline{10}$ flavour representations.

\noindent We add that if the spin of this cluster is 3/2 the only state is:
\begin{eqnarray}\label{8}
       \psi_3 &=& |(q_1 q_2)^{\overline{3}}_1 \rangle \otimes |
       (\overline{q}_3)^{\overline{3}}_{1/2} \rangle \nonumber \\
\end{eqnarray}
It is a pure state in the  $\overline{84} $ of $SU6_{cs}$ and the eigenvalue of
$H_{\mathrm{CM}}$ is  $-\frac{8}{3}\,{\it C_{12}}+\frac{2}{3} ({\it
C_{13}} +{\it
C_{23}})$.

\section{ The $(qq\overline{q})^{15}_{s}$ ``triquark"}

For this last colour configuration, basis states for a total spin $1/2$ cluster
can be chosen as follows:
\begin{eqnarray}\label{8}
       \mu_1 &=& |(q_1 q_2)^{{6}}_1 \rangle \otimes |
       (\overline{q}_3)^{\overline{3}}_{1/2} \rangle \nonumber \\
       \mu_2 &=& |(q_1 q_2)^{{6}}_0 \rangle \otimes |
       (\overline{q}_3)^{\overline{3}}_{1/2} \rangle \nonumber
\end{eqnarray}
\noindent leading to the Hamiltonian:

\begin{equation}\label{9}
H_{\mathrm{CM}}=-\left[ \begin {array}{cc} \frac{4}{3}\ ({\it
C_{12}} - {\it C_{13}} - {\it
C_{23}}) & - \sqrt {3} \left( {\it C_{13}}-{\it C_{23}} \right)
\\\noalign{\medskip}
    -\sqrt {3} \left( {\it C_{13}}-{\it C_{23}} \right) & -4\,
     {\it C_{12}} \end {array} \right]
\end{equation}\\

Finally, if the total spin is $3/2$, then  only one state survives , and
the corresponding eigenvalue
of $H_{\mathrm{CM}} $ reads $\frac{4}{3}\,{\it C_{12}}+\frac{2}{3}
({\it C_{13}} +{\it
C_{23}})$.

\section{Conclusion}
\noindent We have presented a detailed computation of the
colour-magnetic Hamiltonian
for s-wave triquark clusters in case of flavour symmetry breaking.
Such a study appears to us of some importance in this time where the possible
existence of exotics hadrons is again considered. Indeed, the two main
properties of such eventual states being their low masses  and their narrow
widths, a precise  evaluation by the theory of their masses becomes
essential. Let us emphasize once more that the expressions which are
provided can be used as well as for the simple isospin breaking as for
   triquark cluster containing quarks simultaneously light and  heavy
   quarks.\\
   As we have noted before \cite{hogso}, it is awkward to explain the
   low mass of todays
   experimental exotic pentaquark signals with colourmagnetic interactions
   only. But whatever the dynamics can be, the colourmagnetic interaction must
    play a role.\\
\noindent One may wonder to what extend the above results could be use to
easily determine the expression of $H_{\mathrm{CM}}$ with flavour
symmetry breaking for clusters with more than three quarks. Let us
answer in part to this question by considering the case of colour
singlet, s-wave  $(qq\overline{q}\overline{q})$
states: then, using a simple argument, one can deduce that a simple
substitution in the
matrix $ H_{\mathrm{CM}}$  of Eq.(21) provides the desired result.

\section{Acknowledgments}
We thank the editors of Fizika for the honour of inviting us to give this contribution
to the memory of the late professor Dubravko Tadi\'{c} .

\end{document}